\documentclass[reprint,amsmath,amssymb,aps,pre]{revtex4-2}
\usepackage{graphicx}
\usepackage{color}
\usepackage{amsmath}
\usepackage{amssymb}
\usepackage{amsthm}
\usepackage{extarrows}
\usepackage{booktabs}
\usepackage{float}
\usepackage[colorlinks,urlcolor=blue,linkcolor=blue,anchorcolor=blue,citecolor=blue]{hyperref}

\definecolor{DB}{RGB}{0, 24, 191}
\usepackage{mathptmx}

\begin{document}


\title{Emergence of Gamma-Type Upward-Phase Statistics in the Collatz Map: An Effective Poisson Process Mechanism}

\author{Weicheng Fu$^{1,2,3}$}
\email{fuweicheng@tsnu.edu.cn}
\author{Xiaobin Liu$^{1}$}
\email{liuxb@tsnu.edu.cn}
\author{Yisen Wang$^{3,4}$}
\email{wys@lzu.edu.cn}

\affiliation{
$^{1}$ \quad Department of Physics, Tianshui Normal University, Tianshui 741001, Gansu, China\\
$^{2}$ \quad Key Laboratory of Atomic and Molecular Physics $\&$ Functional Material of Gansu Province, College of Physics and Electronic Engineering, Northwest Normal University, Lanzhou 730070, China\\
$^{3}$ \quad Lanzhou Center for Theoretical Physics, Lanzhou University, Lanzhou 730000, Gansu, China\\
$^4$ \quad Key Laboratory for Quantum Theory and Applications of the Ministry of Education, Lanzhou University, Lanzhou, Gansu 730000, China}

\date{\today }


\begin{abstract}

The Collatz map is a simple deterministic transformation whose orbit structure remains highly nontrivial. A recent direction-phase decomposition partitions each orbit into upward and downward steps, and numerical observations indicate that the number of upward phases, $N_{\uparrow}$, follows an approximate Gamma distribution. In this work, we provide a mechanistic explanation for this statistical regularity by modeling the occurrence of upward phases in the odd-compressed, or Syracuse, version of the Collatz map as a homogeneous Poisson process. From the mean-field logarithmic balance and the geometric distribution of $2$-adic valuations, we derive closed-form expressions for the Gamma parameters: the scale parameter $\theta = 2/(2-\log_2 3)^2 \approx 11.61$ is constant, whereas the shape parameter $K$ grows logarithmically with the maximal initial value $X_0=2L+1$. We also analyze the closure conditions for periodic orbits, showing that nontrivial cycles are severely constrained, which supports the plausibility of the statistical framework. Numerical validation for $L$ ranging from $10^5$ to $10^{15}$ confirms the theory with relative errors below $3\%$, and a bias-corrected mean estimate reduces the error to $10^{-3}$--$10^{-2}\%$. These results establish a quantitative link between the arithmetic properties of the Collatz map and Gamma-type statistics, and suggest possible extensions to generalized Collatz-type problems.

\end{abstract}

\maketitle

\section{Introduction}
\label{sec:1}

The Collatz map is a classical problem at the intersection of discrete dynamical systems \cite{1978Crandall,PhysRevLett.49.1801,
Krasikov1989,Abraham1997,wirsching1998dynamical,wirsching2006dynamical,Siegel2024}, elementary number theory \cite{akin20043x+,lagarias2010ultimate,wang2022proof}, and probabilistic modeling \cite{Borovkov2001,Sinai2003}. It is defined by
\begin{equation}\label{eq-3xp1}
X=\begin{cases}
3X+1, & X~\text{is odd};\\
X/2, & X~\text{is even}.
\end{cases}
\end{equation}
The Collatz conjecture asserts that, for every positive integer $X$, repeated iteration  eventually enters the period-three cycle
$4\rightarrow 2\rightarrow 1\rightarrow 4$.
The problem was proposed by L. Collatz in the 1930s \cite{lagarias2010ultimate}. Despite its elementary formulation, the associated orbit structure is highly intricate, and a complete proof remains out of reach. A substantial body of work has investigated the problem from the perspectives of stopping times, total stopping times, modular structures, random models, $2$-adic dynamics, and probabilistic methods \cite{terras1976stopping,Lagarias1985,tao2022almost,Polli_2024}.

In recent years, several noteworthy developments have further advanced the study of the Collatz problem. One important direction is to understand the typical behavior of Collatz orbits through probability theory and random dynamical models \cite{tao2022almost,Polli_2024}. Tao proved that almost all Collatz orbits attain almost bounded values. Although this result does not show that every orbit reaches $1$, it demonstrates that, in the sense of logarithmic density, the overwhelming majority of orbits exhibit a strong tendency toward descent \cite{tao2022almost}. This work refines the long-standing random-walk heuristic for the Collatz problem into a more precise probabilistic framework, suggesting that the deterministic Collatz map may display effective randomness at appropriate scales. Closely related to this viewpoint, random models of Collatz orbits have long served as an important tool for understanding the problem. Kontorovich and Lagarias systematically studied random models for the $3X+1$ and $5X+1$ problems, showing that forward iteration can be approximated by additive random walks, biased random walks, or Markov processes \cite{lagarias2010ultimate}.

Another active direction concerns large-scale computational verification. Although computational evidence cannot replace a rigorous proof, it provides important support for understanding the statistical structure of orbits and for excluding counterexamples within large finite ranges \cite{Barina2021,Barina2025}. Barina recently extended the verification of the Collatz conjecture to $X<2^{71}$, using improved algorithms and parallel computation, and discussed the role of GPU and distributed computing in accelerating this task \cite{Barina2025}. Such computations show that no nontrivial cycle or divergent orbit appears within an extremely large range of initial values.

Beyond stopping-time analysis and computational verification, recent work has also attempted to characterize the structure of Collatz orbits from the perspective of nonlinear dynamical systems. In Ref.~\cite{FuWang2026}, Collatz orbits were decomposed into upward and downward phases, leading to a direction-phase decomposition and a family of recursive functions parameterized by the number of upward phases $N_{\uparrow}$. A key numerical observation in that work is that the statistics of odd integers classified by the number of upward phases are well described by a Gamma-type distribution. This suggests that the number of local growth events in Collatz orbits is not merely an irregular count, but may instead exhibit a stable statistical law.

The present work is motivated by the following question: what mechanism gives rise to Gamma-type statistics in the number of upward phases of Collatz orbits? To answer this question, we model the occurrence of upward phases in the odd-compressed, or Syracuse, version of the Collatz map as a homogeneous Poisson process. Combining the mean-field logarithmic balance with the geometric distribution of the $2$-adic valuations, we obtain closed-form estimates for the Gamma parameters and explain why the scale parameter is approximately constant whereas the shape parameter grows logarithmically with the maximal initial value. Large-scale numerical experiments are then performed to validate the proposed Poisson process mechanism and the resulting Gamma distribution approximation. Beyond its implications for the statistical analysis of Collatz dynamics, this framework also provides a pedagogically useful example for undergraduate and graduate courses in nonlinear dynamics, number theory, computational physics, and computational mathematics, illustrating how deterministic arithmetic rules can be effectively described through probabilistic and statistical models.

The remainder of this paper is organized as follows. Section~\ref{sec:2} introduces the Collatz map and defines the direction-phase decomposition, including the numbers of upward and downward phases. Section~\ref{sec:3} develops a probabilistic approximation for the odd-compressed Collatz dynamics and derives the Gamma distribution approximation for the statistics of $N_{\uparrow}$, together with theoretical estimates for the fitting parameters and the mean value. Section~\ref{sec:4} discusses exact closure conditions for possible periodic orbits and clarifies the role of finite-size corrections in the double-logarithmic asymptotic picture. Section~\ref{sec:5} presents numerical results that test the theoretical predictions over a wide range of $L$. Finally, Section~\ref{sec:6} summarizes the main conclusions and discusses the scope and limitations of the proposed approximation.

\section{The Collatz map and definition of direction phases}\label{sec:2}

For brevity, the map (\ref{eq-3xp1}) is abbreviated as
\begin{equation}\label{eq-model-m}
  X_n = F(X_{n-1}) = F^n(X_0),
\end{equation}
where $ n, X_n \in \mathbb{Z}^+ $. The Collatz conjecture asserts
\begin{equation}\label{eq-C-conjuecture}
\lim_{n \to \infty} F^n(X_0) = \{4,2,1\} \quad \text{for } \forall X_0 \in \mathbb{Z}^+.
\end{equation}
To formalize the dynamics, we introduce ``direction phases'' \cite{Wang2000}:
\begin{equation}\label{eq-phase}
\begin{cases}
P_{\uparrow}(n) = 1, &\text{if} ~X_{n+1} > X_n\quad \text{(up phase)}; \\
P_{\downarrow}(n)= -1, &\text{if}~ X_{n+1} < X_n\quad \text{(down phase)}.
\end{cases}
\end{equation}

Notably, $N_{\uparrow}$ and $N_{\downarrow}$ correspond to the counts of odd and even terms in the sequence. The total iterations satisfy
\begin{equation}\label{eq-Nsum}
N = N_{\uparrow} + N_{\downarrow}, \quad \text{with } F^N(X_0) = 1,
\end{equation}
where $N_{\downarrow}$ is parameterized as a function of $N_{\uparrow}$ as follows
\begin{equation}\label{eq-Nsum2}
N_{\downarrow} =  \left\lfloor N_{\uparrow}\log_2\left(3+\frac{1}{X_0}\right)+\log_2(X_0)\right\rceil,
\end{equation}
here $\lfloor x \rceil$ denotes rounding $x$ to the nearest integer \cite{FuWang2026}. Clearly, for a given $X_0$, the total number of iterations is completely determined under the condition that $N_{\uparrow}$ is known. Numerical experiments show that, within a specified range, the statistical distribution of $N_{\uparrow}$ corresponding to odd $X_0$ approximately follows a Gamma distribution \cite{FuWang2026}, which will be theoretically analyzed and estimated below.

\section{Poisson-process mechanism for Gamma-type upward-phase statistics}\label{sec:3}

The standard Collatz map consists of an odd step, $3X+1$, and an even step, $X/2$. Since $3X+1$ is always even for any odd integer $X$, an odd step is necessarily followed by one or more successive divisions by $2$. It is therefore natural to combine these consecutive even steps and consider the odd-only accelerated map acting on the set of positive odd integers:
\begin{equation}
X_{n+1}
=
\frac{3X_n+1}{2^{h_n}},
\qquad
h_n=\nu_2(3X_n+1),
\end{equation}
where $\nu_2(m)$ denotes the $2$-adic valuation of $m$, i.e., the exponent of the highest power of $2$ dividing $m$. This map is also referred to as the reduced Collatz function, the accelerated Collatz function, the odd-only Collatz map, or the Syracuse function \cite{tao2022almost}. In this compressed representation, each step from $X_n$ to $X_{n+1}$ consists of one upward operation, $X_n\mapsto 3X_n+1$, followed by $h_n$ downward divisions by $2$. Hence, for an orbit segment containing $M$ compressed steps, the number of upward phases is $N_{\uparrow}=M$, while the total number of downward phases is
\begin{equation}
N_{\downarrow}=\sum_{n=0}^{M-1}h_n.
\end{equation}
Thus, the number of iterations of the odd-compressed map is naturally identified with the number of upward phases.

\begin{figure}[t]
  \centering
  \includegraphics[width=1\columnwidth]{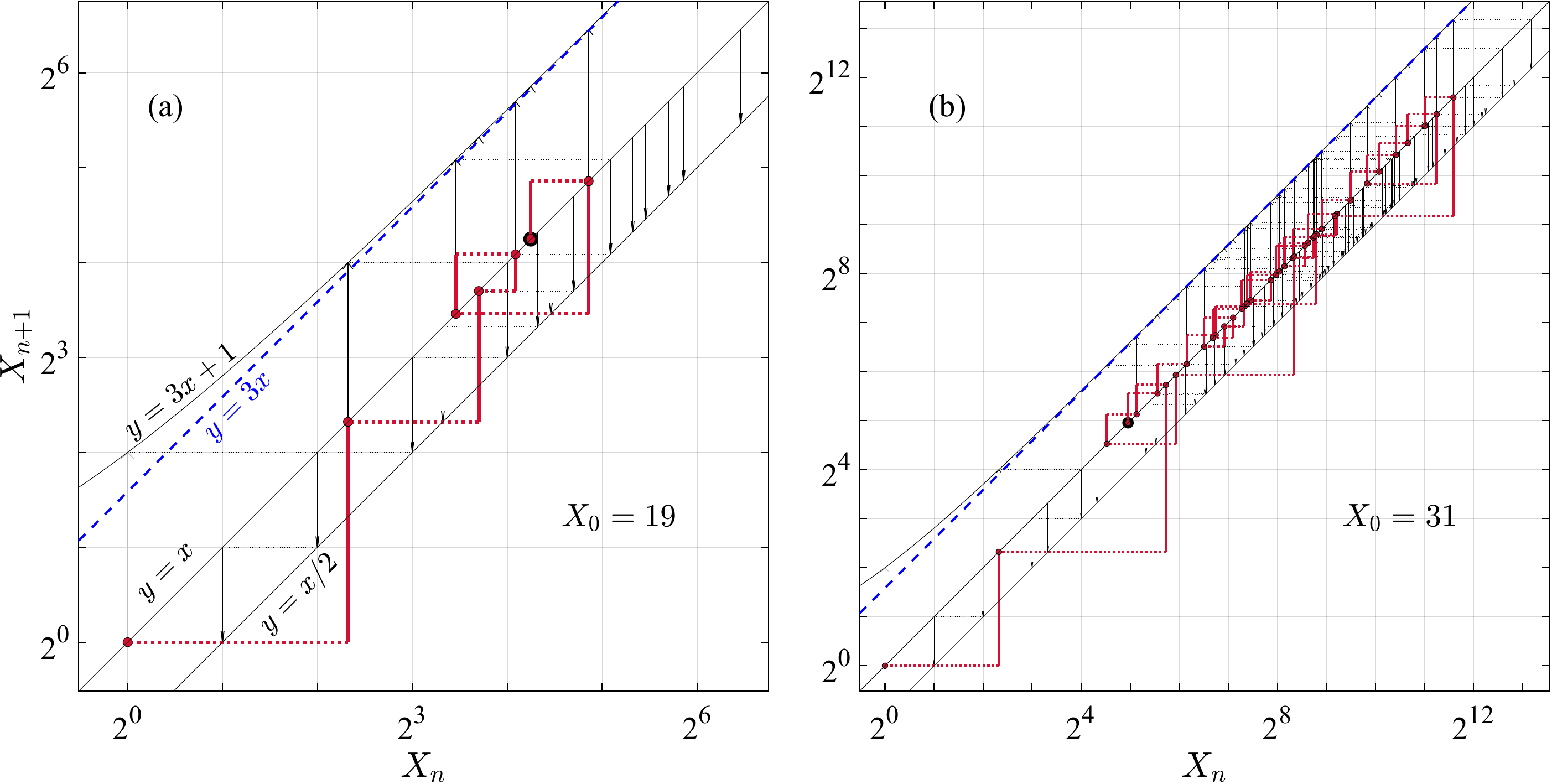}
  \caption{Cobweb plot of the orbits with $X_0=19$ (a) and $X_0=31$ (b), respectively. The red lines are the iterative trajectories of the odd-compressed Collatz map.}\label{fig-Cmap}
\end{figure}

Figures~\ref{fig-Cmap}(a) and \ref{fig-Cmap}(b) compare the iteration trajectories of the odd-compressed map, indicated by the red lines, with the cobweb plots of the original Collatz map for the initial values $X_0=19$ and $X_0=31$, respectively. For these two initial values, the corresponding numbers of upward and downward phases are $(N_{\uparrow},N_{\downarrow})=(6,14)$ and $(39,67)$, respectively. The trajectory generated by the odd-compressed map consists entirely of odd integers, as illustrated by the red dots in Figure~\ref{fig-Cmap}. In the $\log_2$--$\log_2$ coordinate system, the grid lines correspond to powers of two. Therefore, except for the fixed point $X=1=2^0$, no point along the odd-compressed trajectory can coincide with these power-of-two grid lines.

We now introduce an effective statistical approximation. In the odd-compressed Collatz dynamics, each compressed iteration corresponds to one upward phase. We assume that the occurrence of these upward phases can be approximated by a homogeneous Poisson process mechanism. Under this viewpoint, the Gamma distribution arises naturally as the distribution of an accumulated waiting time in a Poisson process. We therefore use the Gamma family as a continuous approximation to the statistics of $N_{\uparrow}$, i.e.,
\begin{equation}\label{eq-rho}
\rho(N_\uparrow)
=\frac{1}{\Gamma(K)\theta^K}
N_\uparrow^{K-1}
\exp\left(-\frac{N_\uparrow}{\theta}\right),
\end{equation}
where $K$ and $\theta$ are the shape and scale parameters, respectively.

The remaining task is to estimate $K$ and $\theta$ from the arithmetic structure of the odd-compressed Collatz map. Taking logarithms of the compressed map yields
\begin{equation}\label{eq-LogXn}
\log_2 X_{n+1}=\log_2(3X_n+1)-h_n.
\end{equation}
For sufficiently large $X_n$, we use the approximation
\begin{equation}\label{eq-LogXn2}
\log_2(3X_n+1)\simeq \log_2(3X_n)
=\log_2 X_n+\log_2 3.
\end{equation}
Substituting this approximation into Eq.~(\ref{eq-LogXn}) gives
\begin{equation}
\log_2 X_{n+1}
\simeq
\log_2 X_n+\log_2 3-h_n.
\end{equation}
Equivalently, the logarithmic evolution can be written as
\begin{equation}
\log_2 X_{n+1}
\simeq
\log_2 X_n-\xi_n,
\end{equation}
where
\begin{equation}
\xi_n=h_n-\log_2 3
\end{equation}
denotes the single-step logarithmic decrease. Since $h_n$ has mean value larger than $\log_2 3$, the average value of $\xi_n$ is positive, corresponding to a net logarithmic contraction on average.

For odd integers that are approximately uniformly distributed among residue classes modulo powers of $2$, the $2$-adic valuation $h_n=\nu_2(3X_n+1)$ may be approximated by a geometric distribution \cite{tao2022almost}, i.e.,
\begin{equation}
\mathbb{P}(h_n=h)
\simeq2^{-h},
\qquad
h=1,2,3,\ldots .
\end{equation}
This approximation gives
\begin{equation}
\mathbb{E}[h_n]=2,
\qquad
\mathrm{Var}(h_n)=2.
\end{equation}
Since $\xi_n=h_n-\log_2 3$, it follows that
\begin{equation}
\mu\equiv\mathbb{E}[\xi_n]
=2-\log_2 3\simeq0.4150375,
\end{equation}
and
\begin{equation}
\sigma^2\equiv\mathrm{Var}(\xi_n)
=\mathrm{Var}(h_n)=2.
\end{equation}

For an initial odd integer $X_0$, reaching the small attracting cycle requires the accumulated logarithmic decrease to be of the order of $\log_2 X_0$. At the mean-field level, this logarithmic balance gives
\begin{equation}
\mu \langle N_\uparrow\rangle
\simeq
\log_2 X_0.
\end{equation}
Taking $X_0=2L+1$ as the representative upper boundary of the sampling interval, we obtain
\begin{equation}\label{eq-av-N}
\begin{aligned}
\langle N_\uparrow\rangle
&\simeq
\frac{\log_2 X_0}{\mu}
=\frac{\log_2 (2L+1)}{2-\log_2 3}
\approx
\frac{1+\log_2 L}{2-\log_2 3}\\
&\approx
2.4094+8.0039\log_{10}L.
\end{aligned}
\end{equation}
This estimate is in good agreement with the numerical results reported in Ref.~\cite{FuWang2026}.

We next estimate the variance of $N_\uparrow$ by error propagation. After $N_\uparrow$ compressed steps, the accumulated logarithmic decrease is
\begin{equation}
S_{N_\uparrow}
=
\sum_{n=1}^{N_\uparrow}\xi_n .
\end{equation}
Within the independent-increment approximation, the variables $\xi_n$ have variance $\sigma^2$, and hence, for a fixed value of $N_\uparrow$,
\begin{equation}
\mathrm{Var}(S_{N_\uparrow})
\simeq
\sigma^2 N_\uparrow .
\end{equation}
At the mean-field level, the accumulated logarithmic decrease is related to the number of upward phases by
\begin{equation}
S_{N_\uparrow}
\simeq
\mu N_\uparrow .
\end{equation}
Thus, a small fluctuation in $S_{N_\uparrow}$ induces a corresponding fluctuation in $N_\uparrow$ according to
\begin{equation}
\Delta N_\uparrow
\simeq
\frac{\Delta S_{N_\uparrow}}{\mu}.
\end{equation}
It follows that
\begin{equation}
\mathrm{Var}(N_\uparrow)
\simeq
\frac{\mathrm{Var}(S_{N_\uparrow})}{\mu^2}
\simeq
\frac{\sigma^2}{\mu^2}\langle N_\uparrow\rangle .
\end{equation}
For a Gamma distribution with shape parameter $K$ and scale parameter $\theta$, one has
\begin{equation}
\mathbb{E}[N_\uparrow]=K\theta,
\qquad
\mathrm{Var}(N_\uparrow)=K\theta^2 .
\end{equation}
Therefore,
\begin{equation}\label{eq-thetaT}
\theta
=
\frac{\mathrm{Var}(N_\uparrow)}
{\mathbb{E}[N_\uparrow]}
\simeq
\frac{\sigma^2}{\mu^2}
=
\frac{2}{(2-\log_2 3)^2}
\simeq
11.6106,
\end{equation}
and
\begin{equation}\label{eq-KT}
\begin{aligned}
K&=\frac{\mathbb{E}[N_\uparrow]^2}
{\mathrm{Var}(N_\uparrow)}
\simeq
\frac{\langle N_\uparrow\rangle}{\theta}
\simeq
\frac{2-\log_2 3}{2}(1+\log_2 L)\\
&\approx
0.2075+0.6894\log_{10}L.
\end{aligned}
\end{equation}
Thus, the theoretical estimate predicts that $\theta$ is independent of $L$, while $K$ grows logarithmically with $L$. In the effective Poisson process interpretation, the constant scale parameter corresponds to a constant inverse rate, supporting the use of a homogeneous Poisson process approximation for the occurrence of upward phases in the odd-compressed Collatz dynamics.

\section{Closure conditions for periodic orbits}
\label{sec:4}

The statistical description developed above implicitly concerns orbits that eventually enter the known cycle. If other nontrivial periodic orbits existed, their long-time behavior would not be described by the same convergent-orbit statistics. It is therefore useful to examine the exact closure conditions for periodic orbits of the accelerated Collatz map.

Suppose that there exists a periodic orbit of the accelerated map on positive odd integers with period $M$, namely,
\begin{equation}
X_0\to X_1\to\cdots\to X_{M-1}\to X_0,
\end{equation}
where $X_M=X_0$. Iterating the accelerated map gives the exact balance condition
\begin{equation}\label{eq-balance}
2^{N_\downarrow}
=\prod_{n=0}^{M-1}\left(3+\frac{1}{X_n}\right),
\end{equation}
where $N_\downarrow=\sum_{n=0}^{M-1}h_n$ is the total number of downward divisions by $2$ along the cycle. For the known cycle $1\to4\to2\to1$, the accelerated map has the fixed point $X_0=1$, corresponding to $M=1$ and $N_\downarrow=2$. In this case, Eq.~(\ref{eq-balance}) reduces to $2^2=3+1$.

Equation~(\ref{eq-balance}) may be rewritten as
\begin{equation}
\frac{2^{N_\downarrow}}{3^M}
=\prod_{n=0}^{M-1}\left(1+\frac{1}{3X_n}\right).
\end{equation}
For any nontrivial positive cycle, all odd elements satisfy $X_n\geq3$. Hence,
\begin{equation}
1<\frac{2^{N_\downarrow}}{3^M}
<\left(\frac{10}{9}\right)^M.
\end{equation}
Taking logarithms, we obtain the necessary condition
\begin{equation}
\log_2 3
<\frac{N_\downarrow}{M}
<\log_2\frac{10}{3}.
\end{equation}
Moreover, if $m=\min\{X_0,X_1,\ldots,X_{M-1}\}$,
then
\begin{equation}\label{eq-N2s}
0<\frac{N_\downarrow}{M}-\log_2 3
<\log_2\left(1+\frac{1}{3m}\right)<\frac{1}{3m\ln2}.
\end{equation}
Thus, any large nontrivial cycle would require the rational number $N_\downarrow/M$ to approximate the irrational number $\log_2 3$ from above with extremely high accuracy.

For example, the numerical verification of the Collatz conjecture up to $X_0<2^{71}$ implies that any possible nontrivial positive cycle must have $m>2^{71}>2.361183\times10^{21}$. Substituting this lower bound into Eq.~(\ref{eq-N2s}) gives
\begin{equation}
\begin{aligned}
0<\frac{N_\downarrow}{M}-\log_2 3
<\frac{1}{3\times2^{71}\ln 2}
<2.0368\times10^{-22}.
\end{aligned}
\end{equation}
Therefore, if one considers a hypothetical sequence of nontrivial cycles with $m\to\infty$, then $N_\downarrow/M$ would have to converge to $\log_2 3$ from above with an error tending to zero.

In the asymptotic approximation where $3X_n+1$ is replaced by $3X_n$, the closure condition would reduce to
\begin{equation}
N_\downarrow=M\log_2 3.
\end{equation}
This equation cannot hold exactly, since $N_\downarrow$ and $M$ are integers whereas $\log_2 3$ is irrational. Hence, in the double-logarithmic asymptotic picture, nontrivial periodic closure is excluded in the limit $m\to\infty$. The constraint becomes increasingly stringent as the minimum element $m$ of the cycle increases.

However, this asymptotic obstruction is not a proof of the nonexistence of all nontrivial cycles. The finite correction $1/X_n$ in $3X_n+1$ cannot be discarded rigorously in the integer dynamics, because it determines the $2$-adic valuation and hence the parity structure of the subsequent trajectory. Indeed, this correction is precisely what produces the known cycle $1\to4\to2\to1$. Therefore, a complete exclusion of all finite-size corrections would amount to proving the uniqueness of the known Collatz cycle. This illustrates the subtle difference between taking asymptotic limits in the real-valued logarithmic approximation and preserving the exact arithmetic structure of the dynamics on positive integers. For example, in the real-valued asymptotic sense, the approximation $3X_n+1\simeq 3X_n$ is valid as $X_n\to\infty$. In the integer dynamics, however, this approximation is not innocuous: for odd $X_n$, the quantity $3X_n+1$ is even, whereas $3X_n$ remains odd. Thus, replacing $3X_n+1$ by $3X_n$ changes the parity structure and, consequently, the $2$-adic valuation that determines the subsequent divisions by $2$.

\section{Numerical results}\label{sec:5}

\begin{figure}[t]
  \centering
  \includegraphics[width=1\columnwidth]{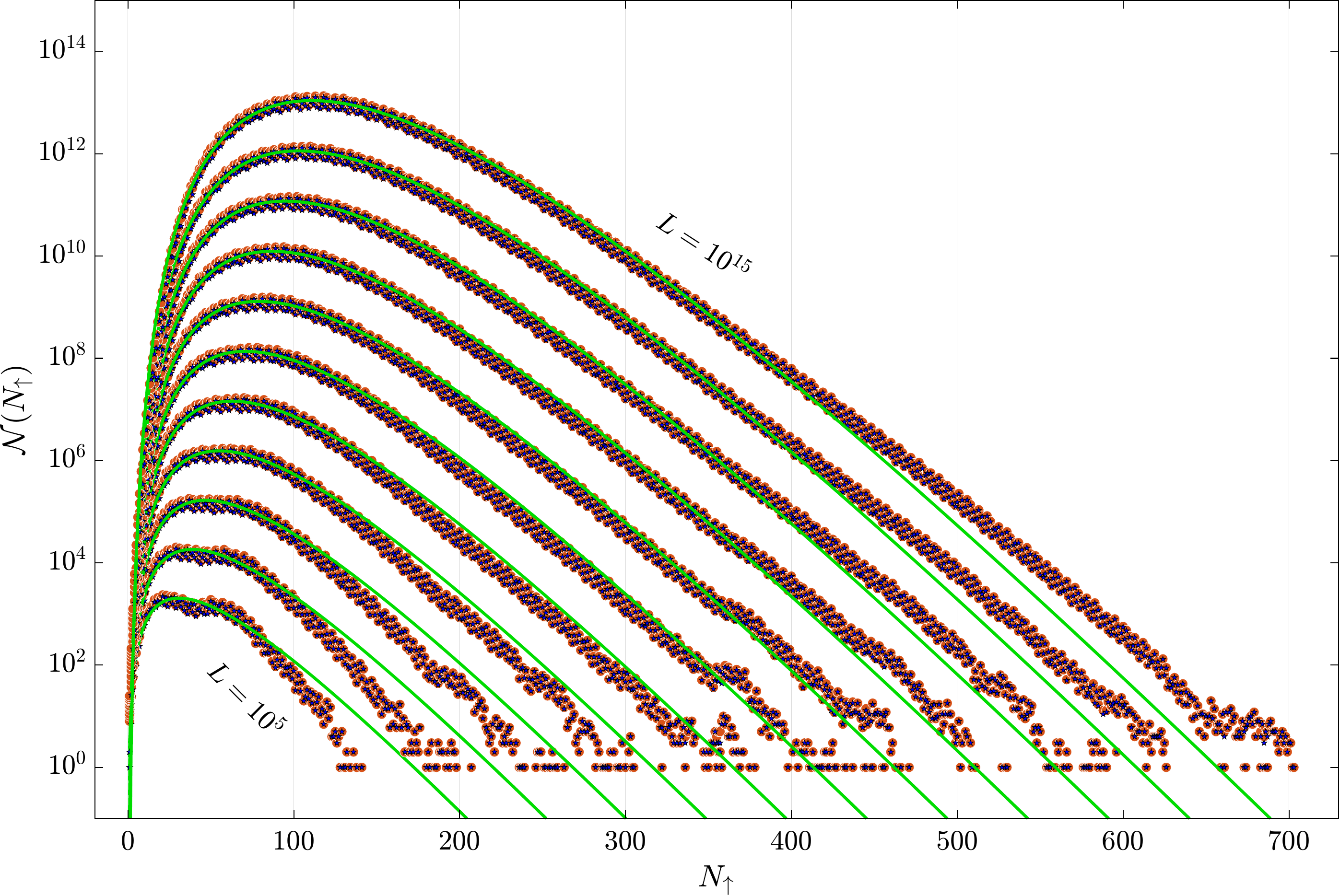}
  \caption{The orange dots denote the counts $\mathcal{N}(N_{\uparrow})$ as a function of $N_{\uparrow}$ for $X_0 \in \{3,5,\ldots,2L+1\}$, plotted on a semilogarithmic scale. The dots from bottom to top correspond to $L=10^5,\ldots,10^{15}$. The blue stars are obtained in the same way, but with $X_0$ restricted to $X_0 \in \{2L\times 10^{-1}+1,\ldots,2L+1\}$. The green solid curves are the theoretical estimate, i.e., $\mathcal{N}(N_{\uparrow})=L\rho(N_{\uparrow})$, see Eq. (\ref{eq-rho}).}\label{fig-Count}
\end{figure}

\begin{figure}
  \centering
  \includegraphics[width=1\columnwidth]{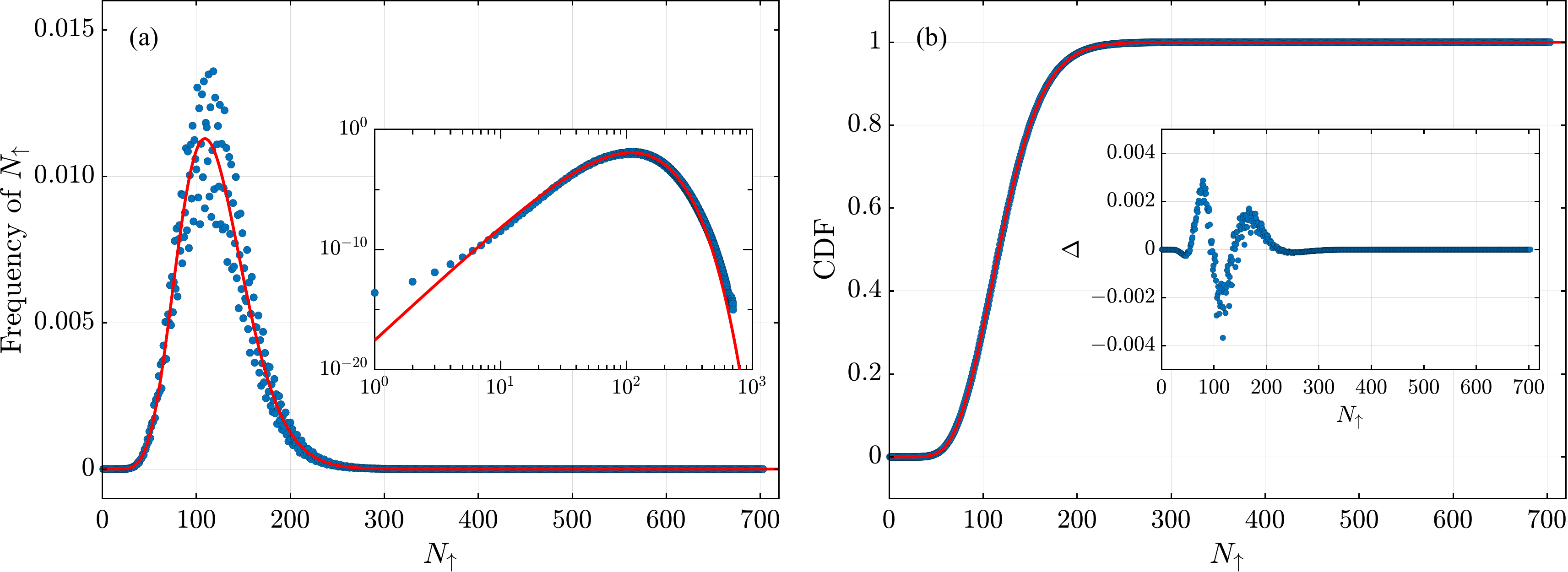}
  \caption{(a) Frequency distribution of $N_{\uparrow}$ as a function of $N_{\uparrow}$ for $X_0 \in \{3,5,\ldots,2L+1\}$ at fixed $L=10^{15}$. The red solid curve denotes the Gamma distribution fit. Inset: the same data plotted on a log-log scale. (b) Empirical cumulative distribution function of the frequency statistics. The solid curve represents the cumulative distribution function of the fitted Gamma distribution. The fitting parameters are $K=10.70924 \pm 0.00621$ and $\theta=11.24522 \pm 0.00661$, with a reduced chi-square of $4.3560117\times10^{-7}$ and an adjusted $R^2$ of $0.9999961$. Inset: residuals between the empirical data and the fitted Gamma distribution, defined as the difference between the empirical and fitted values.}\label{fig-GammaCDF}
\end{figure}

Figure~\ref{fig-Count} shows the statistical counts of $N_{\uparrow}$ for $X_0 \in \{3,5,\ldots,2L+1\}$ and $L=10^5,\ldots,10^{15}$ on a semilogarithmic scale. The orange dots denote the numerical counts, while the blue stars show the corresponding results obtained by restricting $X_0$ to odd values in the range $X_0\in\{0.2L+1,0.2L+3,\ldots,2L+1\}$. The two sets of data nearly overlap, indicating that, for sufficiently large $X_0$, the statistical distribution of $N_{\uparrow}$ is insensitive to the lower cutoff of the sampling interval. The green curves represent the theoretical prediction obtained from the probability density function in Eq.~(\ref{eq-rho}), namely $\mathcal{N}(N_{\uparrow})=L\rho(N_{\uparrow})$, which agrees well with the numerical results.

Figure~\ref{fig-GammaCDF}(a) shows the frequency distribution of $N_{\uparrow}$ as a function of $N_{\uparrow}$ for $X_0 \in \{3,5,\ldots,2L+1\}$ at fixed $L=10^{15}$. The red solid curve denotes the Gamma distribution fit. To better resolve the discrepancy between the numerical data and the fitted curve at small frequencies, the same data are replotted on a double-logarithmic scale in the inset. The fitted curve shows excellent overall agreement with the numerical results. Figure~\ref{fig-GammaCDF}(b) presents the empirical cumulative distribution function constructed from the frequency statistics. Since the cumulative curve is smooth, we fit it using the cumulative distribution function of the Gamma distribution to determine the fitting parameters. The fitted curve in Fig.~\ref{fig-GammaCDF}(a) is then drawn using these parameters. The red solid curve in Fig.~\ref{fig-GammaCDF}(b) denotes the fitted cumulative distribution function. To further illustrate the fitting accuracy, the inset shows the residual $\Delta$ between the empirical and fitted values. The residual satisfies $|\Delta|<4\times10^{-3}$, indicating that the Gamma distribution provides an accurate approximation to the statistics of $N_{\uparrow}$.

\begin{figure}[t]
  \centering
  \includegraphics[width=1\columnwidth]{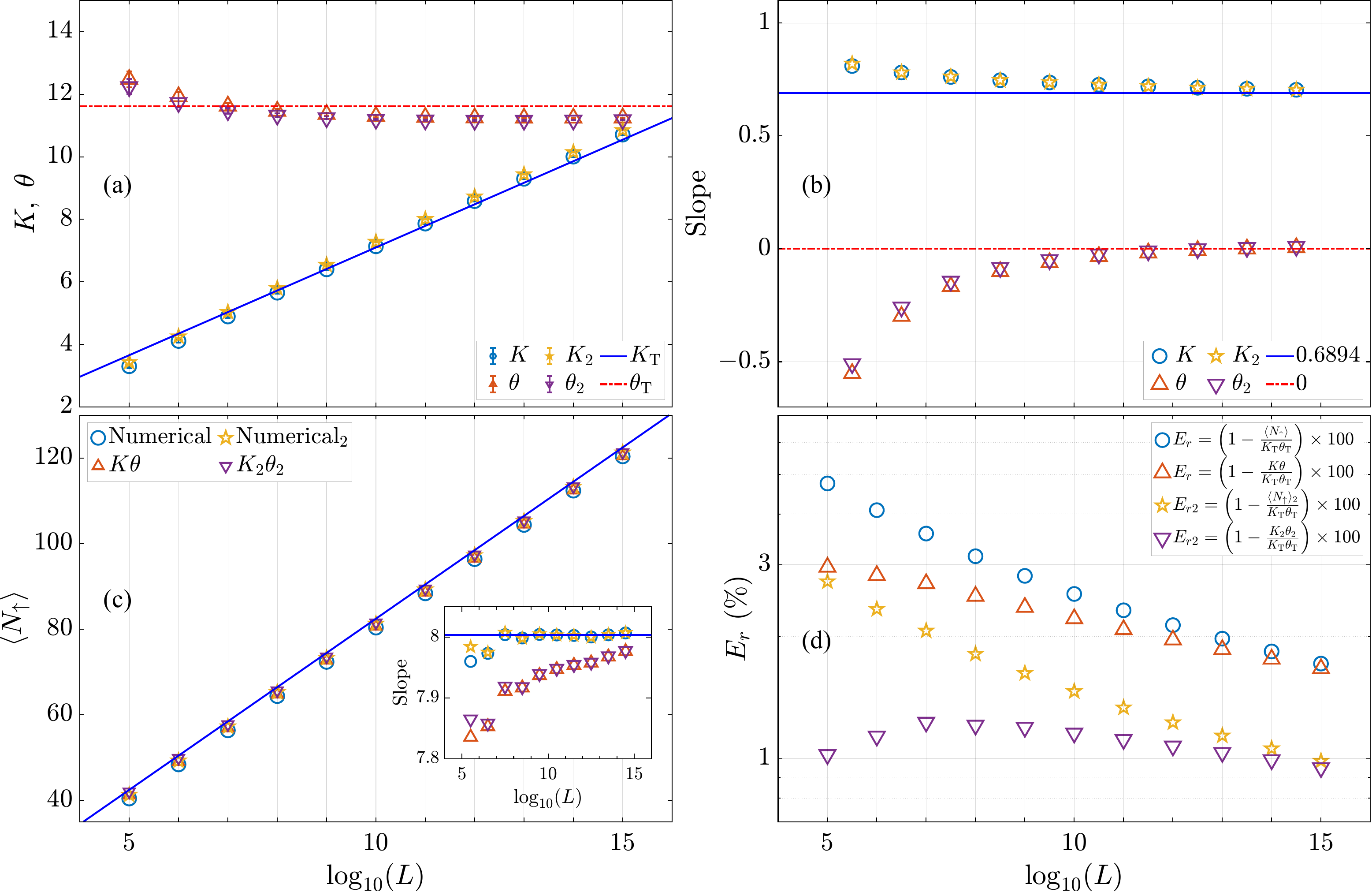}\\
  \caption{(a) Fitting parameters $\theta$ (upward triangles) and $K$ (circles) of the Gamma distribution as functions of $L$, for $X_0\in\{3,5,\ldots,2L+1\}$. The solid and dash-dotted lines denote the theoretical predictions $\theta_{\mathrm{T}}$ [Eq. (\ref{eq-thetaT})] and $K_{\mathrm{T}}$ [Eq. (\ref{eq-KT})], respectively. (b) Local slopes obtained from linear fits between adjacent data points in panel (a) as functions of $L$. The horizontal reference lines indicate the corresponding theoretical values. (c) Statistical mean of $N_{\uparrow}$ (circles) and the mean estimated from the Gamma distribution fit, $K\theta$ (upward triangles), as functions of $L$. The solid line represents the theoretical estimate $K_{\mathrm{T}}\theta_{\mathrm{T}}$ [Eq.~(\ref{eq-av-N})]. Inset: local slopes obtained from linear fits between adjacent data points in the main panel. The horizontal reference line indicates the theoretical value. (d) Relative error between the mean of $N_{\uparrow}$ shown in panel (c) and the corresponding theoretical prediction. Quantities labeled with the subscript $2$ are obtained in the same way, but with $X_0$ restricted to $X_0\in\{0.2L+1,\ldots,2L+1\}$.
}\label{fig-KTheta}
\end{figure}

To quantitatively characterize the dependence of the Gamma approximation on the sampling range of $X_0$, Fig.~\ref{fig-KTheta}(a) shows the fitted Gamma parameters $K$ and $\theta$ as functions of $L$. As $L$ increases, the fitted value of $\theta$ gradually approaches an approximately constant value, $\theta\simeq 11.245$, which is slightly smaller than the theoretical prediction $\theta_{\mathrm{T}}\simeq 11.61$. The fitted value of $K$ changes from slightly below to slightly above the theoretical prediction, while its overall logarithmic growth remains consistent with the theory. Quantities without subscripts in the legend are obtained from the statistics for $X_0 \in \{3,5,\ldots,2L+1\}$, whereas quantities with the subscript $2$ are obtained from the restricted range $X_0 \in \{0.2L+1,\ldots,2L+1\}$. The difference between the two data sets is small, indicating that the lower cutoff of $X_0$ does not affect the qualitative behavior of the distribution.

Figure~\ref{fig-KTheta}(b) further examines this dependence by plotting the local slopes obtained from linear fits between adjacent data points for $K$ and $\theta$. The slope of $\theta$ approaches zero as $L$ increases, consistent with an asymptotically constant scale parameter. In contrast, the slope of $K$ converges toward its theoretical value, as indicated by the horizontal reference line.

Figure~\ref{fig-KTheta}(c) shows the mean value of $N_{\uparrow}$ as a function of $L$. The mean estimated from the Gamma distribution fit agrees closely with the directly computed statistical mean, whereas the theoretical estimate is slightly larger. The inset shows the corresponding local slopes obtained from adjacent-point linear fits. The slope of the statistical mean is nearly identical to the theoretical prediction. Although the slope obtained from the fitted mean is slightly smaller, it gradually approaches the theoretical value $8.0039$ as $L$ increases.

Figure~\ref{fig-KTheta}(d) shows the relative errors of the statistical mean and the fitted mean with respect to the theoretical prediction, i.e., $\langle N_{\uparrow}\rangle=K_{\mathrm{T}}\theta_{\mathrm{T}}$. The relative error remains below $3\%$ over the full range of $L$. When $X_0$ is restricted to the range from $0.2L+1$ to $2L+1$, the relative error is further reduced to approximately $2\%$. Overall, the error decreases with increasing $L$, consistent with the large-$X_0$ approximation used in Eq.~(\ref{eq-LogXn2}). Thus, larger values of $X_0$ lead to smaller approximation errors in the theoretical derivation.

\begin{figure}[t]
  \centering
  \includegraphics[width=1\columnwidth]{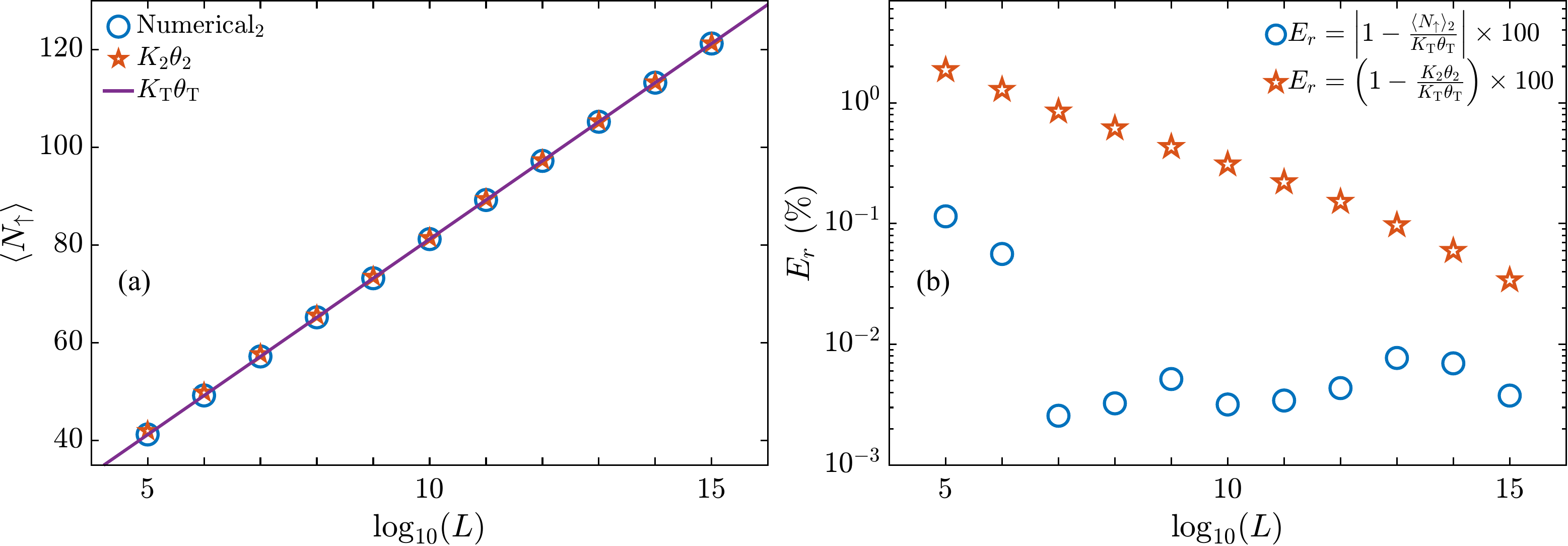}
  \caption{Same as panels (c) and (d) of Fig. \ref{fig-KTheta}, but with the theoretical estimate replaced by Eq. (\ref{eq-av-N2}), where $X_0\in\{0.2L+1,0.2L+3,\dots,2L+1\}.$ }\label{fig-KTheta3}
\end{figure}

Figures~\ref{fig-KTheta}(c) and \ref{fig-KTheta}(d) show that the original theoretical prediction slightly overestimates the mean value of $N_{\uparrow}$, while accurately capturing its scaling slope. When $X_0$ is restricted to the interval $X_0\in\{0.2L+1,0.2L+3,\ldots,2L+1\}$, the numerical results agree better with the theoretical prediction. This indicates that using the upper boundary $X_0=2L+1$ in Eq.~(\ref{eq-av-N}) as the representative logarithmic scale of the whole sampling interval leads to a systematic overestimation for smaller values of $X_0$. This bias decreases as $X_0$ approaches $2L+1$. To reduce this bias, we replace $\log_2(2L)$ in Eq.~(\ref{eq-av-N}) by the arithmetic mean of the logarithmic small interval $[\log_2 L,\log_2(2L)]$, namely $1/2+\log_2 L$. The corrected theoretical estimate is therefore
\begin{equation}\label{eq-av-N2}
\langle N_{\uparrow}\rangle
\simeq
\frac{1/2+\log_2 L}{2-\log_2 3}
\approx
1.2047+8.0039\log_{10}L.
\end{equation}

Figure~\ref{fig-KTheta3}(a) compares this corrected prediction with the numerical results, showing good agreement. Figure~\ref{fig-KTheta3}(b) further displays the relative error of the directly computed statistical mean as a function of $L$. With the corrected theory, the relative error is reduced to the order of $10^{-3}\%$--$10^{-2}\%$. In addition, the relative error of the mean estimated from the fitted Gamma parameters decreases and falls below $0.1\%$ at $L=10^{15}$.

\section{Summary and discussion}
\label{sec:6}

In summary, we have provided a mechanistic explanation for the emergence of Gamma-type statistics in the direction-phase structure of Collatz orbits. By decomposing the Collatz dynamics into upward and downward phases and using the odd-compressed representation, we modeled the occurrence of upward phases through an effective homogeneous Poisson process mechanism. Together with the geometric approximation for the $2$-adic valuations $h_n$, this framework leads naturally to a Gamma-family approximation for the statistics of $N_\uparrow$. The corresponding Gamma parameters can be estimated analytically from the arithmetic structure of the map. The scale parameter is predicted to be $\theta_{\mathrm{T}}
=\frac{\sigma^2}{\mu^2}
=\frac{2}{(2-\log_2 3)^2}
\simeq 11.61$. In contrast, the shape parameter $K$ grows logarithmically with $L$, consistent with the mean-field logarithmic balance between the accumulated decrease and $\log_2 X_0$. A corrected estimate for the mean value of $N_\uparrow$, obtained by replacing the upper-boundary approximation with the average logarithmic scale of the sampling interval, further reduces the relative error to the order of $10^{-3}\%$--$10^{-2}\%$.

We also examined the closure conditions for possible periodic orbits of the accelerated Collatz map. The exact balance condition $
2^{N_\downarrow}
=\prod_{n=0}^{M-1}\left(3+{1}/{X_n}\right)$
implies that any nontrivial large cycle would require $N_\downarrow/M$ to approximate $\log_2 3$ from above with an accuracy better than $1/(3m\ln2)$, where $m$ is the minimum element of the cycle. As $m \to \infty$, this condition becomes arbitrarily stringent. In the asymptotic approximation where $3X_n+1$ is replaced by $3X_n$, the closure condition reduces to $N_\downarrow=M\log_2 3$, which is impossible because $\log_2 3$ is irrational. This provides an asymptotic obstruction to large nontrivial cycles, although it does not constitute a rigorous exclusion of all finite-size corrections.

Several questions remain open. The effective Poisson process assumption is strongly supported by the numerical results, but a rigorous derivation from the deterministic Collatz dynamics is still lacking. In particular, it would be important to quantify correlations among successive values of $\nu_2(3X_n+1)$ and to determine whether a suitable limit theorem can be established for the odd-compressed map. Another direction is to extend the direction-phase framework to generalized Collatz-type maps, such as $aX+b$ problems, in order to clarify whether the observed Gamma-type statistics are specific to the classical $3X+1$ map or reflect a broader feature of parity-driven integer dynamics.

Beyond its implications for the statistical analysis of Collatz dynamics, the present framework also provides a useful pedagogical example. For undergraduate and graduate teaching, it provides a compact example showing how an elementary deterministic rule can generate highly nontrivial behavior and how number-theoretic structure, discrete dynamics, probabilistic modeling, and numerical evidence can interact. In this sense, the Collatz problem is not only a classical topic in elementary number theory, but also an instructive model for illustrating how deterministic arithmetic dynamics may admit effective stochastic descriptions, while numerical evidence can motivate but not replace rigorous mathematical analysis.

\begin{acknowledgments}
This research was funded by the National Natural Science Foundation of China (Grants
No. 12465010, No. 12247106, and No.12575040). W. Fu was also supported by the Gansu Province Long-yuan Youth Talent Project, the Fei-tian Scholars Project of Gansu Province, the Leading Talent Project of Tianshui City, the Innovation Fund from the Department of Education of Gansu Province (Grant No. 2023A-106), the Open Project Program of the Key Laboratory of Atomic and Molecular Physics $\&$ Functional Material of Gansu Province (Grant No. 6016-202404), and Young Faculty Achievement Award (Grant No. PYJ-01252291) $\&$ Graduate Program Construction Project (Grant No. TKXM2601) of Tianshui Normal University.
\end{acknowledgments}


%

\end{document}